\documentclass[aps,pra,twocolumn,showpacs,amssymb,amsmath,amsfonts,superscriptaddress,floatfix,nofootinbib,groupedaddress]{revtex4-2}
\usepackage{graphicx}
\usepackage{mathtools}
\usepackage{perpage} 
\usepackage{amsmath,amssymb}
\MakePerPage{footnote}
\usepackage{color}
\usepackage{braket}
\usepackage{subfigure}

\begin{document}


\title{Continuous quantum clock with high precision and long recurrence time}


\author{Mehdi Ramezani, Morteza Nikaeen, Alireza Bahrampour}
\affiliation{Department of Physics, Sharif University of Technology, Tehran 14588, Iran}


\date{\today}

\begin{abstract}
Continuous clocks, i.e. the clocks that measure time in a continuous manner, are regarded as an essential component of sensing technology. Precision and recurrence time are two basic features of continuous clocks. In this paper, in the framework of quantum estimation theory various models for continuous quantum clocks are proposed, where all tools of quantum estimation theory are employed to seek the characteristics of clocks with high precision and long recurrence time. Then, in a resource-based approach, the performance of the proposed models is compared. It is shown that quantum clocks based on $n$ two-qubits system not only can have better precision than quantum clocks based on $2n$ one-qubit system but also support long recurrence time. Finally, it is shown that while employing the number of $n$ entangled qubits improves the precision of clocks by a factor of $1/\sqrt n $, it inevitably worsens the recurrence time of the clock.
\end{abstract}

\maketitle

\section{Introduction}\label{introduction}
Although the time in its abstract form emerges as a dilemma in quantum mechanics \cite{muga2007time}, it has a simple operational definition. From an operational point of view, rather than considering the time as universal coordinate $t$  in space-time, time is considered as a dynamical variable of a physical system in space-time \cite{hilgevoord2002time}. This is just where the concept of clocks comes into play.

The basic idea behind the operation of any clock is that, as the position variable $q$ of a point particle resembles the space coordinate $x$, physical systems exist that have a dynamical variable that resembles in the same way the time coordinate $t$. From this perspective, the estimation of the time has a direct correspondence to the dynamics description of physical systems that at the fundamental level is given by the laws of quantum mechanics.

To present the motivation of our proposed quantum clock, it is constructive to distinguish between two kinds of clocks, i.e. discrete and continuous clocks. Discrete clocks, like pointer clocks, employ the dynamical systems which pass through a succession of recordable states at constant time intervals, therefore can only record the precise time at discrete intervals and are inaccurate within the intervals. Continuous clocks like hourglass or sundial, however, record the continuous dynamical evolution of the system, capable of measuring the time in a continuous manner. Therefore, while with discrete clocks we are concerned with both accuracy (i.e. tick intervals) and precision (i.e. ticking on true time), in continuous clocks we are concerned only with precision (i.e. the resolution of clocks).

The recurrence time (RT) of a clock, i.e. the maximum time interval in which a clock can measure, is another characteristic of a clock we are concerned with. For discrete clocks which are periodic systems that work with the so called tiks (i.e. the recorded states of the system in regular intervals), one can count the number of tiks to measure the arbitrary long time intervals. However, the characterization of RT is of serious concern for continuous clocks, because there is nothing similar to tiks in these clocks.

It is also instructive to distinguish between two types of clocks mechanism, deterministic vs. probabilistic mechanism. In the deterministic mechanism there is a deterministic relation between the time, $t$ and dynamical variable that represents the time, say $\chi (t)$. To introduce the concept of probabilistic mechanism consider the dynamical variable $X(t)$ of a quantum system with Hamiltonian $H$, where $[H,X] \ne 0$. In this system, although the evolution of $X(t)$ as a quantum dynamical variable is deterministic, its classical description in the Hamiltonian eigenbasis is restricted to quantum measurement with probabilistic outcomes. Therefore, in the so-called probabilistic mechanism, the correspondence is between the time and probability distribution of measurement outcomes, say $P(X|t)$.

A Quantum clock is a quantum system that its dynamical variable $X(t)$ lies in the Hilbert space of the system. There have been proposed various models for a quantum clock \cite{salecker1997quantum},\cite{peres1980measurement},\cite{woods2019autonomous},\cite{buvzek1999optimal},\cite{erker2017autonomous}. The common point of most of these approaches is that the quantum state of a system swipes some specified states $\left| {{\phi _1}} \right\rangle ,\left| {{\phi _2}} \right\rangle ,...,\left| {{\phi _n}} \right\rangle$, respectively in equal time intervals. Therefore, they are considered as discrete quantum clocks. Continuous clocks can be regarded as essential elements of sensors and therefore are on high demand in one of the main components of today’s technology, i.e. metrology and sensing. In this paper we employ the approach of quantum estimation theory (QET) to propose a continuous quantum clock with optimal characteristics. In this regard, the quantum clock is modeled as an encoding-decoding mechanism, where the time is the parameter that is coherently encoded into qubit systems by an optimal Hamiltonian and subsequently decoded via an optimal projective measurement, corresponding to optimal encoding and decoding operators, respectively. Two basic elements of a continuous clock are precision and RT of the clock. Therefore our main problem is to propose a continuous quantum clock with high precision and long RT. Since the proposed clock is based on the probabilistic mechanism, the problem naturally is introduced in the context of QET. Therefore, we should tackle the problem with all tools of QET.

In the first step, we show that, given an ensemble of $n$ qubit systems, one can improve the resolution of the quantum clock by proper choice of the driving Hamiltonian and measurement basis up to Cramer-Rao bound. In this way, the shot noise due to the quantum measurements is the only inevitable source of the noise that determines the ultimate precision of the clock. However, obtaining the time estimator for quantum clock based on ensemble of $n$ qubit systems, shows that there is a trade-off between the precision and  RT of the proposed clock, a problem imposed by the quantum recurrence theorem \cite{bocchieri1957quantum}. In the second step, we show that this problem can be resolved by using two-qubits system instead of one-qubit system. Obtaining the time estimator for quantum clocks based on the dynamics of coherent two-qubit systems shows that one can introduce coarse and fine estimators of the time through slowly and rapidly varying parts of the dynamics, respectively. The frequency of the slowly varying part of the dynamics determines the RT of the clock and can be set to arbitrary small values corresponding to large values of the RT. Instead, this is the frequency value of the rapidly varying part of the dynamics that determines the precision of the quantum clock and independent of the former can be set to arbitrary large values. In this way, the characteristics of a quantum clock with high precision and long RT is obtained.

But this is not quite the whole story. QET gives us another powerful tools that can improve the resolution of the estimators up to Heisenberg limit \cite{holland1993interferometric}, \cite{giovannetti2011advances}, \cite{giovannetti2006quantum}. Actually, the theory suggests that by exploiting more strong resources, i.e. the number of $n$ entanglement resources, the resolution of the estimators scales as $1/n$ that improves the shot noise limit by a factor of $1/\sqrt n $. But a closer investigation indicates that this improvement is solely due to the increasing the evolution frequency of the system that in turns reduces the RT of the clock. Thus, in this scenario there exists a trade-off between the two quantities, that is, while entanglement improves the precision of the quantum clock, it also worsens the RT of the clock.

The proposed models employ a resource-based approach. In this regard, the paper is concluded with a discussion about the comparisons between the performance of the introduced models in terms of the number of used resources. Especially, it is shown that the quantum clocks based on $n$ non-entangled two-qubits systems not only support the clocks with long RT but also its precision outperforms the quantum clocks based on $2n$ one-qubit systems.

\section{Parameter Estimation Theory}\label{PE}
Suppose that outcomes of an experiment is described by a random variable \textbf{X}, where the probability distribution corresponding to the the outcomes of the experiment depends on the parameter $\theta$, i.e. $P(X = x) := {P_x}(\theta )$. Let $\textbf{x} = \{ {x_1},{x_2},{x_3},...,{x_n}\}$ be the random outcomes of the experiment in $n$ trials. An estimator of a parameter $\theta$ is a function $\hat \theta (\textbf{x})$
that assigns a value for $\theta$ from the outcomes of the experiment. The estimator ${\hat \theta }$ is an \textit{unbiased estimator} if $E(\hat \theta ) = \theta$, where $E$ denotes the average of the random variable $\hat \theta$. The theoretical limitation on the resolution of an 
unbiased estimator is given by the Cramer-Rao inequality \cite{cramer2016mathematical}:
\begin{equation}\label{Cramer-Rao}
\Delta \theta  \ge \frac{1}{\sqrt{{nF[P_x(\theta)]}}},
\end{equation}
where $\Delta \theta$, the estimated error, is the standard deviation of the estimator
\begin{equation}\label{variance}
\Delta \theta  = \sqrt{E({\theta ^2}) - E{(\theta )^2}},
\end{equation}
and $F[P_x(\theta)]$ is the so-called Fisher information \cite{cramer2016mathematical}, given by
\begin{equation}\label{Fisherdefine}
F[P_x(\theta)]: = \sum\limits_x {{P_x}(\theta )} {[{\partial _\theta }\log({P_x}(\theta ))]^2}.
\end{equation}

For a quantum system with density matrix $\varrho(\theta)$, subjected to the protective measurement $\{\Pi_x\}$, the probability distribution $P_x(\theta)$ is given by
\begin{equation}
    P_x(\theta) = \text{Tr}[\varrho(\theta)\Pi_x].
\end{equation}

If the parameter $\theta$ is encoded via an observable $A$ according to the following equation
\begin{equation}
    \varrho(\theta) = \exp{(-i\theta A)}\varrho_{0}\exp{(+i\theta A)},
\end{equation}
the quantum Fisher information $F_{\rm {Q}}[\varrho(\theta) ,A]$ attributed to the quantum state $\varrho(\theta)$ and encoder operator $A$ is defined as follows \cite{paris2009quantum}
\begin{equation}
    F_{\rm {Q}}[\varrho(\theta) ,A]=2\sum _{k,l}{\frac {(\lambda _{k}-\lambda _{l})^{2}}{(\lambda _{k}+\lambda _{l})}}\vert \langle k\vert A\vert l\rangle \vert ^{2},
\end{equation}
where $\lambda _{k}$ and $\vert k\rangle$ are the eigenvalues and eigenvectors of the density matrix $\varrho(\theta)$, respectively.

It can be shown \cite{paris2009quantum} that quantum Fisher information is equal to the maximum Fisher information introduced in Eq.(\ref{Fisherdefine}) where the maximization is taken over all possible measurement bases, i.e.
\begin{equation}
    F_{\rm {Q}}[\varrho(\theta) ,A] = \max_{\{\Pi_x\}} F[P_x(\theta)].
\end{equation}

Therefore, in quantum scenarios the optimal estimators are those saturating the inequality given in Eq.(\ref{Cramer-Rao}), where $F[P_x(\theta)]$ is replaced by the quantum version of Fisher information.  

\section{Precise Quantum Clock Based on one-qubit system}
The key idea behind the operation of any typical clock, whether be a classical or quantum system, is to find a one-to-one correspondence between the time and evolution of a dynamical variable of the system. However, the precision of the clock is determined by the nature of the clock system and the underlying laws for describing it. Therefore, the ultimate resolution in measuring time is determined by the laws of quantum mechanics. Although the evolution of a quantum system is deterministic, its classical description required for reading the time is probabilistic. From this perspective, the characteristics of a quantum clock with optimal precision can be determined in the context of statistical approaches. In this section, we show how QET can be employed to design a clock with optimal precision.  

\subsection{The model description}

\begin{figure}[t]
	\centering
	\includegraphics[scale=0.5]{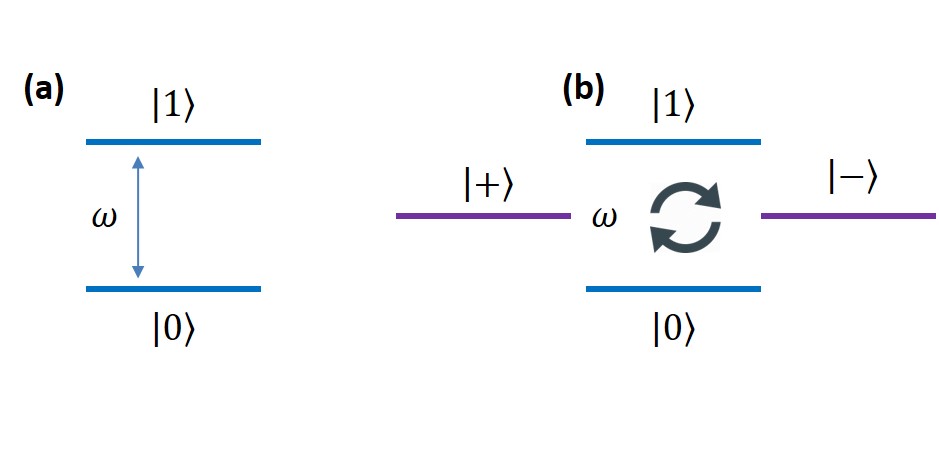}
	\caption{(a) Energy levels of a system with Hamiltonian $H = \frac{\omega }{2}{\sigma _z}$ is considered as a qubit system, where the states $\left| 0 \right\rangle$ and $\left| 1 \right\rangle$ are eigenstates of the Pauli matrix ${\sigma _z}$. (b) When the system initially is prepared in the state $\left| + \right\rangle$, it starts to oscillate between the states $\left| + \right\rangle$ and $\left| - \right\rangle$ with frequency $ \omega$, where the states $\left| \pm  \right\rangle$ are eigenstates of the Pauli matrix ${\sigma _x}$.This dynamics is used to estimation of the time with optimal precision. }\label{optdynamics}
\end{figure}

\begin{figure}[t]
	\centering
	\includegraphics[scale=0.5]{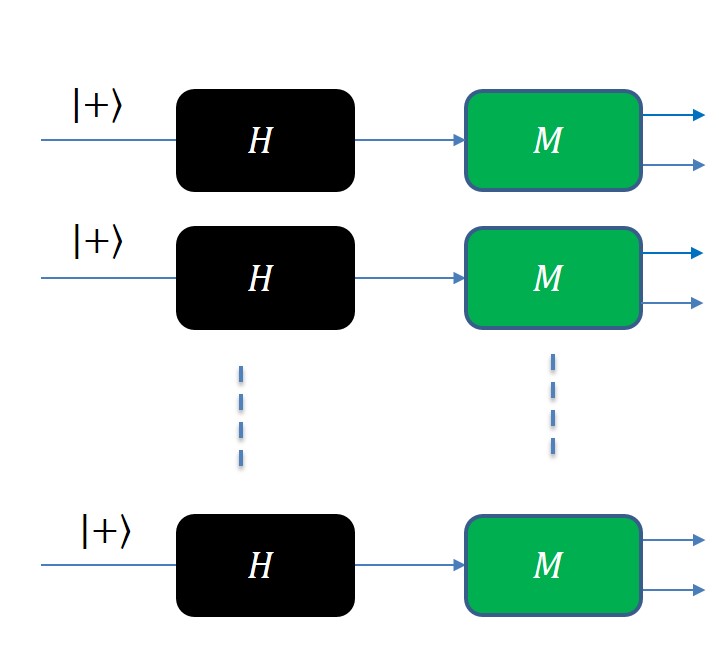}
	\caption{Schematic model of quantum clock based on dynamics of $n$ one-qubit system. All qubits are initially prepared in the  state $\left|  +  \right\rangle$ and then evolve with Hamiltonian $H = \frac{\omega }{2}{\sigma _z}$. At finite time $t$ a measurement in the basis $\left\{ {\left|  \pm  \right\rangle } \right\}$ is carried out on each qubit. From the statistical of the measurement outputs the time $t$ can be estimated, where the minimum error of the estimation scales as $1/(\sqrt n \omega)$. }\label{scenario1}
\end{figure}

Consider a qubit system that in its abstract form corresponds to the Hilbert space $\{ \text{span}(\left| + \right\rangle ,\left| - \right\rangle )\}$ and assume the system initially is prepared in the state $\left| {\psi (0)} \right\rangle  = \left| + \right\rangle$. Suppose that the Hamiltonian eigenstates of the system are expended in terms of the basis $\left\{ {\left|  +  \right\rangle ,\left|  -  \right\rangle } \right\}$ as follows
\begin{gather}\label{states}
\left| {{\varphi _0}} \right\rangle  = a\left| + \right\rangle  - b\left| - \right\rangle, \\
\left| {{\varphi _1}} \right\rangle  = c\left| + \right\rangle  + d\left| - \right\rangle, 
\end{gather}
where, $\left| {{\varphi _0}} \right\rangle$ and $\left| {{\varphi _0}} \right\rangle$ are eigenstates corresponding to energy levels $E_0$ and $E_1$, respectively and the quantities $a, b, c$ and $d$ are real numbers satisfying the orthogonality and normalization conditions according to the following relations
\begin{gather}\label{normalize}
{a^2} + {b^2} = 1,
{c^2} + {d^2} = 1,
ac - bd = 0.
\end{gather}

The Schrodinger equation then determines the evolution of the system in any given time $t$ as follows
\begin{multline}\label{thyoft}
\left| {\psi (t)} \right\rangle  = \big(\frac{{ad}}{{ad + bc}}{e^{ - i{E_0}t}} + \frac{{bc}}{{ad + bc}}{e^{ - i{E_1}t}}\big)\left| + \right\rangle\\
  + \big(\frac{{ - bd}}{{ad + bc}}{e^{ - i{E_0}t}} + \frac{{bd}}{{ad + bc}}{e^{ - i{E_1}t}}\big)\left| - \right\rangle .
\end{multline}

Form Eq.(\ref{thyoft}) it is evident that performing a measurement with projective operators ${\Pi _+} = \left| + \right\rangle \left\langle + \right|$ and ${{\Pi _-} = \left| - \right\rangle \left\langle - \right|}$ on the system at any time $t$ leads to the following probabilities associated to the measurement outcomes ${\Pi _+}$ and ${\Pi _-}$, respectively:
\begin{align}\label{p0}
    {P_+}(t) &= 1 - \chi {\sin ^2}\big(\frac{{\omega t}}{2}\big),\\\label{p1}
    {P_-}(t) &= \chi {\sin ^2}\big(\frac{{\omega t}}{2}\big),
\end{align}
where, $\omega  = {E_1} - {E_0}$ and $\chi  = \frac{{2abcd}}{{{{(ad + bc)}^2}}}$ with $0 \le \chi  \le 1$.

Now, considering the time-dependent probability distribution given in Eq.(\ref{p0}) and Eq.(\ref{p1}), we are faced with a standard problem in QET, where the parameter to be estimated is time. Following the approach presented in Sec.\ref{PE}, the Fisher information corresponding to the probability distribution $\{ {P_+}(t),{P_-}(t)\}$ is given by
\begin{multline}\label{Fisher}
F[P_i(t)] = \sum\limits_{i = +,-} {{P_i}(t){[{\partial _t }\ln [{P_i}(t)]]}^2} \\= \frac{{{\chi ^2}{\omega ^2}{{\sin }^2}(\omega t)}}{{\chi (1 - 2\chi )(1 - \cos (\omega t)) + {\chi ^2}{{\sin }^2}(\omega t)}}.
\end{multline}

According to the Cramer-Rao inequality, to enhance the precision of estimating time through the measurement outputs, the Fisher information corresponding to the outputs statistics must be maximized. Therefore, optimization over the precision of the quantum clock is equivalent to finding a Hamiltonian that maximize the Fisher information given in Eq.(\ref{Fisher}). This is equivalent to fix the Hamiltonian and find a measurement basis that maximizes the Fisher information. A simple investigation of Eq.(\ref{Fisher}) shows that the Fisher information reaches its maximum value when
\begin{equation}\label{chi}
    \chi  = 1,
\end{equation}
 which implies that 
\begin{equation}\label{F(t)}
    F[P_i(t)] = {\omega ^2}.
\end{equation}

A possible solution for the variables ${a,b,c,d}$ that satisfies Eq.(\ref{normalize}) and Eq.(\ref{chi}) is 
\begin{equation}\label{abcd}
    a = b = c = d = \frac{1}{{\sqrt 2 }}.
\end{equation}

If we consider the states $\left\{ {\left|  +  \right\rangle ,\left|  -  \right\rangle } \right\} $ as eigenstates of the Pauli matrix ${\sigma _x}$, then the states introduced in Eq.(\ref{states}) with the coefficients given by Eq.(\ref{abcd}) are eigenstates of the Pauli matrix ${\sigma _z}$. Thus, the optimal Hamiltonian is of the following form
\begin{equation}\label{optimalHaliltonian}
    H = \frac{\omega }{2}{\sigma _z} = 
    \left( 
    {
    \begin{array}{*{20}{c}}
        {\frac{\omega }{2}}&0\\
        0&{\frac{{ - \omega }}{2}}
    \end{array}} 
    \right).
\end{equation}

Therefore, a measurement process described by the projective operators $\{{\Pi _+},{\Pi _-}\}$ on a qubit system prepared in the initial state $\left| {\psi (0)} \right\rangle  = \left| + \right\rangle$ whose dynamics is driven by the Hamiltonian given by Eq.(\ref{optimalHaliltonian}) maximizes the Fisher information associated to the measurement outputs (see Fig.\ref{optdynamics}). In this way, when $n$ probes are used to estimation the time, the resolution of the estimator obeys the following inequality
\begin{equation}\label{1qubitinequality}
\Delta t \ge \frac{1}{{\sqrt n \omega }}.
\end{equation}

Fig.(\ref{scenario1}) represents the schematic model of the quantum clock based on dynamics of $n$ one-qubit system.

Having characterized the model of quantum clock with optimum precision, in the following section we seek the best estimator of the time, i.e. an operational way to read the time with minimal error through a measurement over an ensemble of $n$ similar qubit systems with mentioned characteristics.

\subsection{The Estimator of the time - Operational way for reading the time}
\begin{figure*}[t]
	\centering
	\includegraphics[scale=0.4]{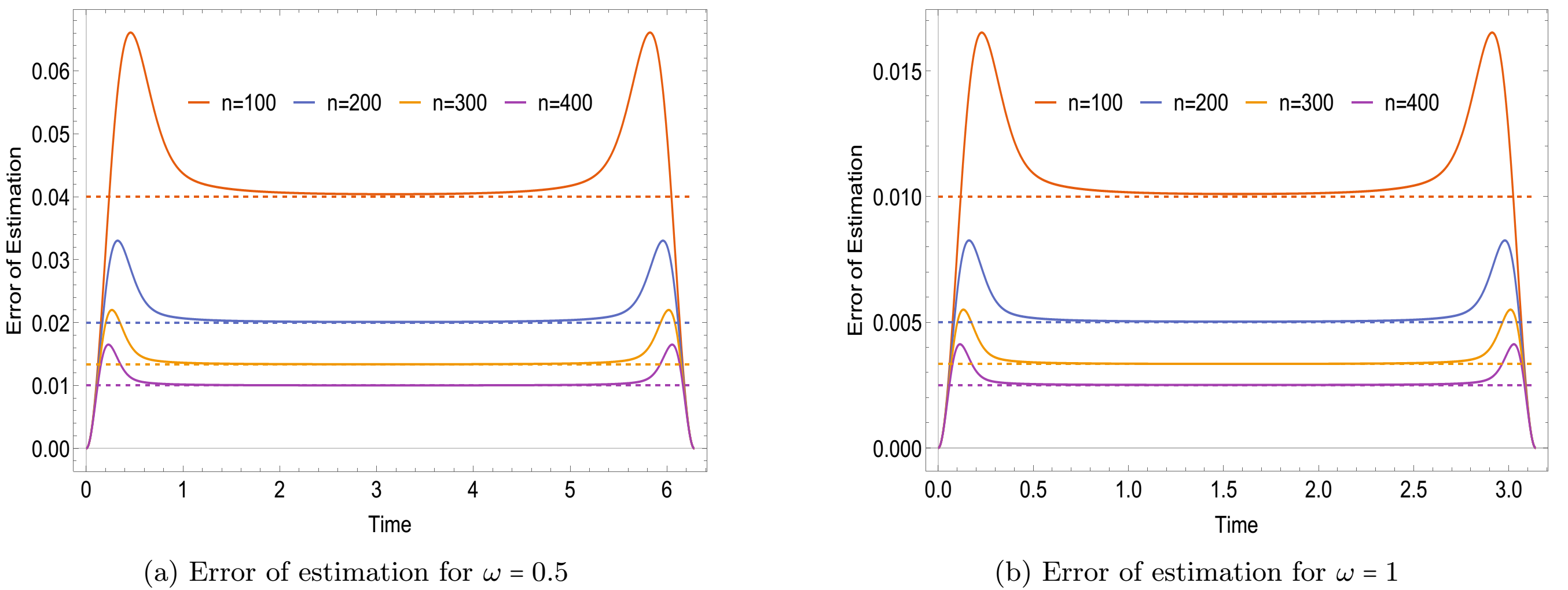}
	\caption{The error of estimation versus time for different values of $n$ and $\omega$. The dashed lines show the minimum error of unbiased
		estimators, i.e. $\frac{1}{{\sqrt{nF[P_{x}(t)]}}}$}\label{ErrorPlot}
\end{figure*}

In this subsection we briefly describe how one can find the best estimator for measuring time. Consider a system of $n$ non-interacting qubits in the initial state
\begin{equation}\label{rho0}
\varrho (0) = {(\left| + \right\rangle \left\langle + \right|)^{ \otimes n}},
\end{equation}
all of which have a Hamiltonian of the form of Eq.(\ref{optimalHaliltonian}). The state of the system at time $t$ is given by
\begin{equation}\label{rhot}
\varrho (t) = {(\left| \psi (t) \right\rangle \left\langle \psi (t) \right|)^{ \otimes n}},
\end{equation}
where 
\begin{equation}\label{thyt}
\left| {\psi (t)} \right\rangle  = \cos (\frac{{\omega t}}{2})\left| + \right\rangle  + \sin (\frac{{\omega t}}{2})\left| - \right\rangle.
\end{equation}

Let $M = \{ {\Pi _1},{\Pi _2},...,{\Pi _n}\}$ be a projective measurement that counts the number of post-measured qubits in the state $\left| - \right\rangle$:
\begin{equation}\label{pik}
{\Pi _k} = \sum\limits_{{i_1} < {i_2} < .. < {i_k}} {\left| {{\varphi _k}({i_1},{i_2},...,{i_k})} \right\rangle } \left\langle {{\varphi _k}({i_1},{i_2},...,{i_k})} \right|,
\end{equation}
where ${\left| {{\varphi _k}({i_1},{i_2},...,{i_k})} \right\rangle }$ is a state in which the $i_1^{th},i_2^{th},...,$ and the $i_k^{th}$ qubits are in the state $\left| - \right\rangle$ and all other qubits are in the state $\left| + \right\rangle$:

\begin{equation}\label{phiparameterization}
    \resizebox{.88\hsize}{!}{$
    \ket{\varphi_{k}({i_1},{i_2},...,{i_k})} = 
    \ket{\underbrace +_{{1^{st}}},\underbrace +_{{2^{nd}}},...,\underbrace -_{i_1^{th}},...,\underbrace -_{i_2^{th}},...,\underbrace 
	-_{i_k^{th}},...,\underbrace +_{{n^{th}}}}.$
	}
\end{equation}

Performing the measurement $M$ at time $t$ yields the outcome associated with ${\Pi _k}$ with probability:
\begin{equation}\label{pkt}
{P_k}(t) = \left( \begin{array}{l}
n\\
k
\end{array} \right){[{\sin ^2}(\frac{{\omega t}}{2})]^k}{[{\cos ^2}(\frac{{\omega t}}{2})]^{n - k}}.
\end{equation}

In order to find a time estimator using the above probability distribution, we employ maximum likelihood method (MLM). The likelihood function, $L(\theta |\textbf{x})$, indicates how likely we are to observe the outcomes of a measurement for a known parameter $\theta$. As a matter of fact, changing $\theta$ will result in different likelihoods for observing our specific measurement
outcomes $\textbf{x}$. For instance if we find $L({\theta _1}|\textbf{x}) > L({\theta _2}|\textbf{x})$, we know that $\textbf{x}$ is more likely to be observed under parameter
conditions $\theta  = {\theta _1}$ rather than $\theta  = {\theta _2}$. This property of likelihood function will lead us to find the best estimator.

Using Eq.(\ref{pkt}), we can write our likelihood function as follows
\begin{equation}\label{likelihood}
L(t|k) = \left( \begin{array}{l}
n\\
k
\end{array} \right){[{\sin ^2}(\frac{{\omega t}}{2})]^k}{[{\cos ^2}(\frac{{\omega t}}{2})]^{n - k}},
\end{equation}
where the unknown parameter is $t$. In order to find the best estimator for time, we should choose $t$ such that it gives us the best likelihood to observe our measurement outcomes. In the other words, the estimator $\hat t(k)$ will be equal to ${t_{\max }}(k)$ by which the likelihood function given in Eq.(\ref{likelihood}) is maximized, i.e.
\begin{equation}\label{differential}
\frac{{\partial L(t|k)}}{{\partial t}}{|_{t(k) = {t_{\max }}(k)}} = 0.
\end{equation}

Solving the above equation for $t$ will give us the time estimator associated to the outcomes of the measurement $M$:
\begin{equation}\label{estimator}
\hat t(k) = \frac{2}{\omega }{\sin ^{ - 1}}(\sqrt {\frac{k}{n}} ).
\end{equation}

To check the precision of the quantum clock, it is instructive to evaluate the error of estimated time, i.e $\Delta t = \sqrt {\left\langle {{t^2}} \right\rangle  - {{\left\langle t \right\rangle }^2}}$, where
\begin{equation}\label{mean}
\left\langle {{t^n}} \right\rangle  = \sum\limits_{k = 1}^n {{P_k}(t){{\hat t}^n}(k)} ,\ \ \ \ \ \ \  n = 1,2
\end{equation}
 in which ${{P_k}(t)}$ and $\hat t(k)$ are given in Eq.(\ref{pkt}) and Eq.(\ref{estimator}), respectively.
 Fig.(\ref{ErrorPlot}) illustrates the numerical results of $\Delta t$ for different values of $n$ and $\omega$. The results confirm that the lower bound of Eq.(\ref{Cramer-Rao}) is saturated; i.e. the proposed clock has a resolution equal to the optimal amount of $\Delta t = \frac{1}{{\sqrt n \omega }}$. This relation indicates that increasing $\omega $, enhances the resolution of the QC. However, according to Eq.(\ref{estimator}), the RT of the system, that is the maximum time interval in which the QC can measure, scales as ${\omega ^{ - 1}}$. Thus, there is a trade-off between the precision and the RT of the QC, i.e. increasing one, decreases the other. In the next section we show that the dynamics of a two-qubit system can be employed to propose a QC with high precision and long RT.

\section{quantum clock with high precision and long recurrence time}\label{schenario2section}
In addition to the precision of the clock, the maximum time interval in which the clock is able to measure is another characteristic of an optimal clock. Our model of quantum clock, is based on the estimation of time through the results of a measurement performed on a dynamical variable of the quantum system. According to the quantum recurrence theorem \cite{bocchieri1957quantum}, for a quantum system in the initial state $\left| {\psi (0)} \right\rangle$ and for any positive number $\epsilon$, there exists a time T such that
\begin{equation}\label{recurrencetheorem}
\left\| {\left| {\psi (T)} \right\rangle  - \left| {\psi (0)} \right\rangle } \right\| < \epsilon.
\end{equation}
Since, for unique determining of time, a one-to-one correspondence between the time and dynamics of the system is necessary, the recurrence time $T$ restricts the time measuring interval in which the quantum clock operates. In this section we aim to circumvent this difficulty by adding another qubit to the system. It is shown that a two-qubits system can support a quantum clock with optimal precision and long RT.

\subsection{Model Description}

Consider a two-qubits system with the Hamiltonian $H$, which in the basis $\{ \left| {00} \right\rangle ,\left| {01} \right\rangle ,\left| {10} \right\rangle ,\left| {11} \right\rangle \}$ is represented as follows
\begin{equation}\label{twoqubitshamiltonian}
H = \left( {\begin{array}{*{20}{c}}
{\frac{\omega }{2}}&0&0&0\\
0&{\frac{{ - \omega }}{2}}&0&0\\
0&0&{\frac{\Omega }{2}}&0\\
0&0&0&{\frac{{ - \Omega }}{2}}
\end{array}} \right),
\end{equation}
and suppose that the system is prepared in the initial state $\left| {\psi (0)} \right\rangle$, where
\begin{equation}\label{key}
\left| {\psi (0)} \right\rangle  = \left|  +  \right\rangle  \otimes \left| + \right\rangle.
\end{equation}

Using the eigen-states and eigen-vectors of the Hamiltonian H, which are of the following form
\begin{gather}
\left| {{\varphi _1}} \right\rangle  = \left| 0 \right\rangle  \otimes \left|  1  \right\rangle ,\ \ \ \ \ {E_1} = \frac{{ - \omega }}{2}\\
\left| {{\varphi _2}} \right\rangle  = \left| 0 \right\rangle  \otimes \left|  0  \right\rangle ,\ \ \ \ \ {E_2} = \frac{{ + \omega }}{2}\\
\left| {{\varphi _3}} \right\rangle  = \left| 1 \right\rangle  \otimes \left|  1  \right\rangle,\ \ \ \ \ {E_3} = \frac{{ - \Omega }}{2}\\
\left| {{\varphi _4}} \right\rangle  = \left| 1 \right\rangle  \otimes \left|  0  \right\rangle,\ \ \ \ \ {E_4} = \frac{{ + \Omega }}{2}
\end{gather}
and considering the initial state of the system, the time evolution of the system is described by the following state 
\begin{multline}\label{twoqubitsthyt}
\left| {\psi (t)} \right\rangle  = \frac{1}{{\sqrt 2 }}\left| 0 \right\rangle  \otimes [\cos (\frac{{\omega t}}{2})\left|+ \right\rangle  - i\sin (\frac{{\omega t}}{2})\left| - \right\rangle ] \,+\\
\frac{1}{{\sqrt 2 }}\left| 1 \right\rangle  \otimes [\cos (\frac{{\Omega t}}{2})\left| + \right\rangle  - i\sin (\frac{{\Omega t}}{2})\left| - \right\rangle ].
\end{multline}

According to the Eq.(\ref{twoqubitsthyt}), if at time $t$ we carry out the  measurement M with the following projective operators:
\begin{align}
    \Pi_{0+} &= \ket{0}\bra{0} \otimes \ket{+}\bra{+},\nonumber\\
    \Pi_{0-} &= \ket{0}\bra{0} \otimes \ket{-}\bra{-},\nonumber\\
    \Pi_{1+} &= \ket{1}\bra{1} \otimes \ket{+}\bra{+},\nonumber\\
    \Pi_{1-} &= \ket{1}\bra{1} \otimes \ket{-}\bra{-},\label{M1}
\end{align}
then the probability of obtaining outcomes associated with ${\Pi _{0+}},{\Pi _{0-}},{\Pi _{1+}}$ and ${\Pi _{1-}}$ are as follows
\begin{align}
    \nonumber
    P_{0+}(t) &= \frac{1}{2} \cos^2\big(\frac{\omega t}{2}\big),
    &
    P_{0-}(t) &= \frac{1}{2} \sin^2\big(\frac{\omega t}{2}\big),
    \\
    \label{jointprob}
    P_{1+}(t) &= \frac{1}{2} \cos^2\big(\frac{\Omega t}{2}\big),
    &
    P_{1-}(t) &= \frac{1}{2} \sin^2\big(\frac{\Omega t}{2}\big).
\end{align}

\begin{figure}[t]
	\centering
	\includegraphics[scale=0.4]{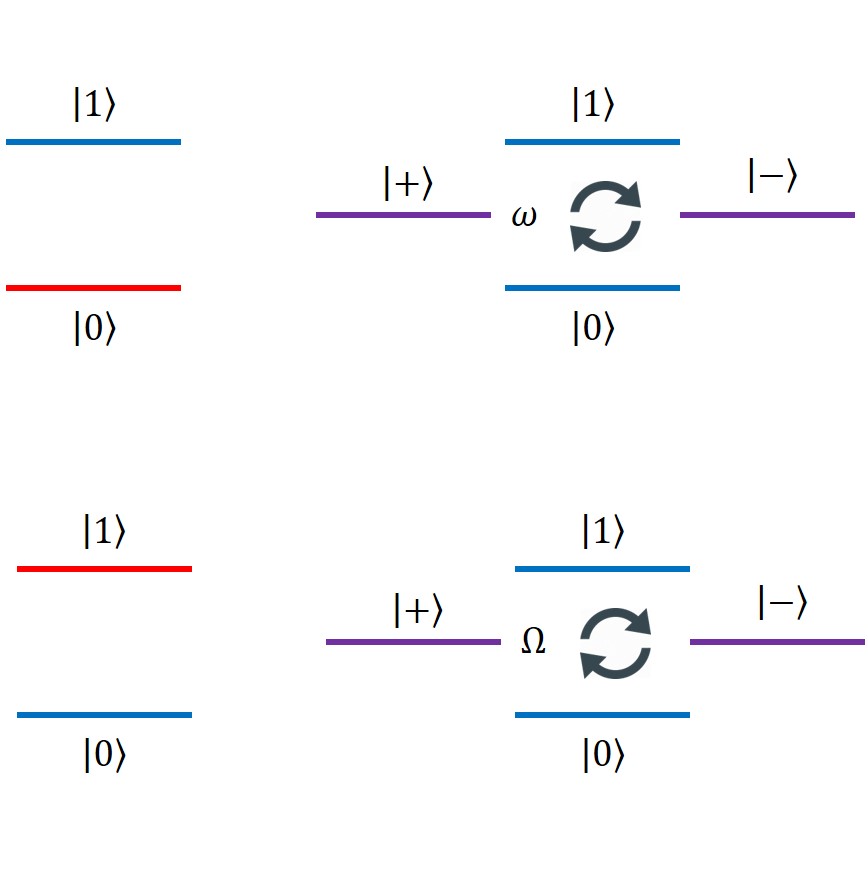}
	\caption{The dynamics of a two-qubits system is employed to propose a quantum clock with high precision and long RT.}\label{twoqubitsdynamics}
\end{figure}

From Eq.(\ref{jointprob}) it is evident when the first qubit is in the state $\left| 0 \right\rangle$ ($\left| 1 \right\rangle$) then the second qubit oscillates between the states $\left| + \right\rangle$ and $\left| - \right\rangle$ with frequency $\omega$ ($\Omega$) (see Fig.(\ref{twoqubitsdynamics})).

The Fisher information of the probability distribution $\{ {P_{0+}},{P_{0-}},{P_{1+}},{P_{1-}}\}$, is given by
\begin{equation}\label{2qbitsFisher}
F[P_x(t)] = \frac{1}{2}({\omega ^2} + {\Omega ^2}).
\end{equation}

Therefore, when $n$ probes are used to estimate time, the resolution of the estimator obeys the following inequality
\begin{equation}\label{secondresolution}
\Delta t \ge \frac{1}{{\sqrt{n(\frac{\omega^2 + \Omega^2}{2})}}}.
\end{equation}

\begin{figure}[t]
	\centering
	\includegraphics[scale=0.4]{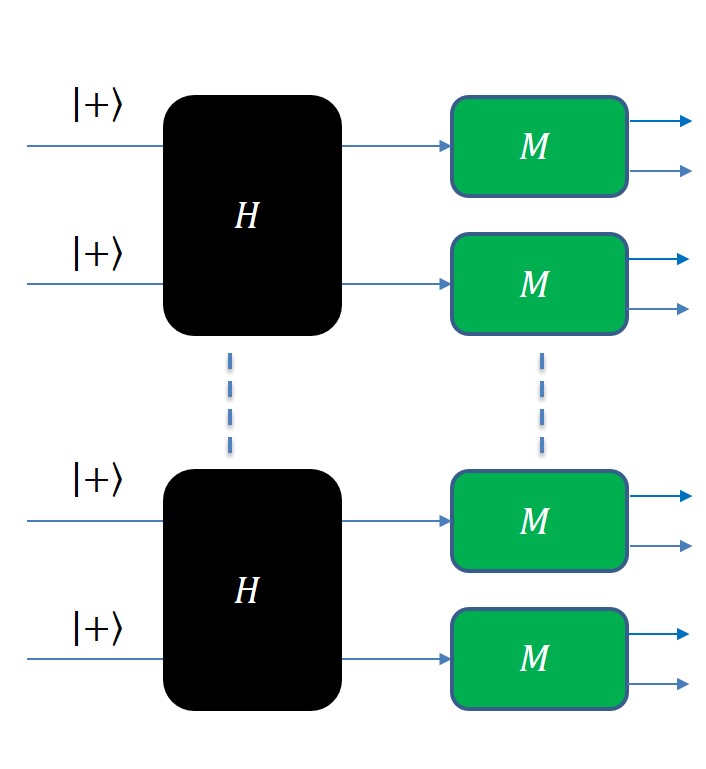}
	\caption{Schematic model of quantum clock based on dynamics of $n$ two-qubits system. All two-qubits system are initially prepared in the  state ${\left|  +  \right\rangle \left|  +  \right\rangle }$ and then evolve with the Hamiltonian given in Eq.(\ref{twoqubitshamiltonian}). At finite time $t$ a measurement in the basis $\left\{ {\left|  \pm  \right\rangle \left|  \pm  \right\rangle } \right\}$ is carried out on each two-qubits system. From the statistical of the measurement outputs the time $t$ can be estimated, where the minimum error of the estimation scales as $\frac{1}{{\sqrt{n(\frac{\omega^2 + \Omega^2}{2})}}}$. In this model the RT scales as $1/{{\min [\omega ,\Omega ]}}$.}\label{scenario2}.
\end{figure}

On the other hand, from Eq.(\ref{twoqubitsthyt}) it is evident that RT of the system scales as
\begin{equation}\label{recurrencetime}
T \sim \frac{1}{{\min [\omega ,\Omega ]}},
\end{equation}
where \textit{min} denotes minimum value.\\
From Eq.(\ref{2qbitsFisher}) and Eq.(\ref{recurrencetime}), we see that decreasing $\omega$ and increasing $\Omega$ at the same time, increases the RT and precision of the quantum clock, respectively. Thus, at the cost of adding one more qubit, one can make a quantum clock with high precision and long RT. Indeed, the RT of the clock can be determined by frequency $\omega$, and the precision of the clock can be determined by frequency $\Omega$ of the system. In this way, as we see in the next section, while the frequency $\omega$ is related to coarse tuning of the quantum clock, the frequency $\Omega$ is responsible for the fine tuning of the proposed quantum clock.
Fig.(\ref{scenario2}) illustrates the schematic representation of the quantum clock based on dynamic of $n$ two-qubits system.

\subsection{Operational way to read the long and precise time}

\begin{figure}[t]
	\centering
	\includegraphics[scale=0.3]{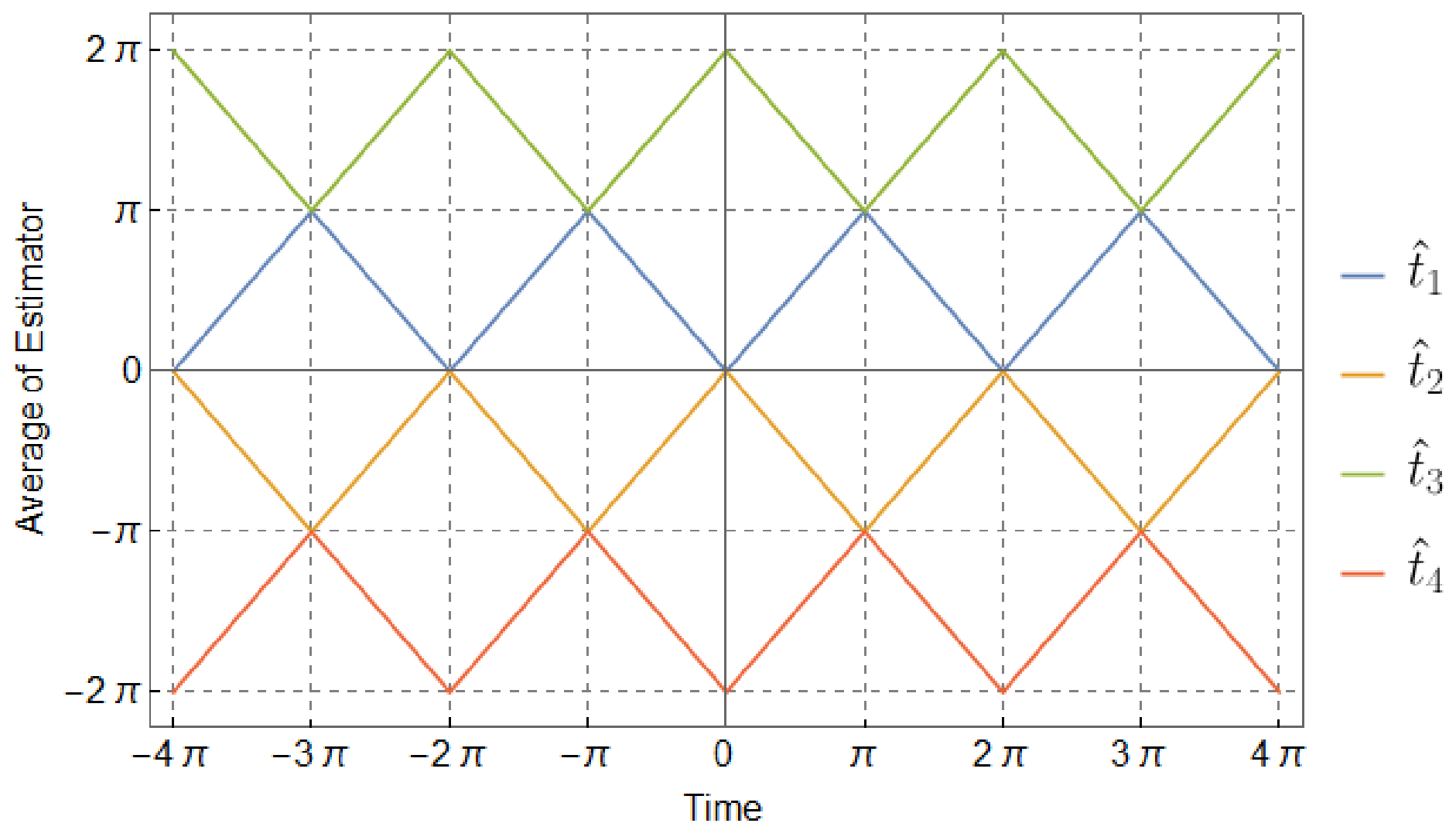}
	\caption{Average of estimators $\hat{t}_{m}(\bf{k})$ with all $l_m = 0$}\label{AvgofE}
\end{figure}

Consider $n$ non-interacting two-qubits systems,where the corresponding state of $2n$-qubits system at time $t$ is represented as follows
\begin{equation}\label{2nthyt}
\varrho (t) = {(\left| {\psi (t)} \right\rangle \left\langle {\psi (t)} \right|)^{ \otimes n}},
\end{equation}
and $\left| {\psi (t)} \right\rangle$ is given by Eq.(\ref{twoqubitsthyt}). If this $2n$-qubits system is subjected to the projective measurement ${M^{ \otimes n}}$ with $M$ introduced in Eq.(\ref{M1}), the probability distribution associated to the outcomes of the measurement ${M^{ \otimes n}}$ at time $t$ is obtained as follows
\begin{multline}\label{jointp}
{P_{\textbf{k}}}(t) = {(\frac{1}{2})^n}\left( {\begin{array}{*{20}{c}}
	n\\
	{{k_1}}
	\end{array}} \right)\left( {\begin{array}{*{20}{c}}
	{n - {k_1}}\\
	{{k_2}}
	\end{array}} \right)\left( {\begin{array}{*{20}{c}}
	{n - {k_1} - {k_2}}\\
	{{k_3}}
	\end{array}}
 \right)\\
 {\cos ^{2{k_1}}}\big(\frac{{\omega t}}{2}\big){\sin ^{2{k_2}}}\big(\frac{{\omega t}}{2}\big){\cos ^{2{k_3}}}\big(\frac{{\Omega t}}{2}\big){\sin ^{2{k_4}}}\big(\frac{{\Omega t}}{2}\big),
\end{multline}
where $\textbf{k} = ({k_1},{k_2},{k_3},{k_4})$ with ${k_1},{k_2},{k_3}$ and ${k_4}$ denote the number of qubits found in the post-measurement states associated to the outcomes ${\Pi _{1-}},{\Pi _{1+}},{\Pi _{0-}}$ and ${\Pi _{0+}}$, respectively; and ${k_1} + {k_2} + {k_3} + {k_4} = n$.
To seek the time estimator associated to this measurement outcomes, once again we employ MLM which was delineated in the previous section. Here, the likelihood function is given by Eq.(\ref{jointp}). Thus by maximizing Eq.(\ref{jointp}) or its logarithm with respect to $t$, one can obtain the desired time estimator. Hence, the time estimator ${\hat t(\textbf{k})}$ is obtained by solving the following equation for $t$:
\begin{equation}\label{eq}
\frac{{\partial \ln ({P_{\textbf{k}}}(t))}}{{\partial t}}{|_{t(\textbf{k}) = \hat t(\textbf{k})}} = 0.
\end{equation}

For simplicity, consider the case in which $\omega = 0.5$ and $\Omega = 1$. For this case, solving the Eq.(\ref{eq}) leads to the following four solutions:
\begin{equation}\label{S1}
\resizebox{1\hsize}{!}{${{\hat t}_1}(\textbf{k}) =  + 4{\tan ^{ - 1}}(\sqrt {\frac{{{k_1} + {k_2} + 4{k_3} + 2{k_4} - \sqrt {{{({k_1} - {k_2})}^2} + 8({k_1} + {k_2} + 2{k_4}){k_3} + 16k_3^2} }}{{2({k_1} + {k_4})}}} ) + 2l_{1}\pi\\$},\\
\end{equation}
\begin{equation}\label{S2}
\resizebox{1\hsize}{!}{${{\hat t}_2}(\textbf{k}) =  - 4{\tan ^{ - 1}}(\sqrt {\frac{{{k_1} + {k_2} + 4{k_3} + 2{k_4} - \sqrt {{{({k_1} - {k_2})}^2} + 8({k_1} + {k_2} + 2{k_4}){k_3} + 16k_3^2} }}{{2({k_1} + {k_4})}}} ) + 2l_{2}\pi\\$},\\
\end{equation}
\begin{equation}\label{S3}
\resizebox{1\hsize}{!}{${{\hat t}_3}(\textbf{k}) =  + 4{\tan ^{ - 1}}(\sqrt {\frac{{{k_1} + {k_2} + 4{k_3} + 2{k_4} + \sqrt {{{({k_1} - {k_2})}^2} + 8({k_1} + {k_2} + 2{k_4}){k_3} + 16k_3^2} }}{{2({k_1} + {k_4})}}} ) + 2l_{3}\pi\\$},\\
\end{equation}
\begin{equation}\label{S4}
\resizebox{1\hsize}{!}{${{\hat t}_4}(\textbf{k}) =  - 4{\tan ^{ - 1}}(\sqrt {\frac{{{k_1} + {k_2} + 4{k_3} + 2{k_4} + \sqrt {{{({k_1} - {k_2})}^2} + 8({k_1} + {k_2} + 2{k_4}){k_3} + 16k_3^2} }}{{2({k_1} + {k_4})}}} ) + 2l_{4}\pi\\$},\\
\end{equation}
where $l_m$, $m \in \{1,2,3,4\}$ are integers. Fig.(\ref{AvgofE}) illustrates the average of estimators for all estimators given in Eq.(\ref{S1})-Eq.(\ref{S4}).

We are only interested in unbiased estimators in positive time domain. From Fig.(\ref{AvgofE}) it is evident that ${\hat t_1}$ is an unbiased estimator in time interval $[0,\pi]$, and ${\hat t_3}$ is an unbiased estimator in time interval $[\pi,2\pi]$. In this way, for measurement outcomes taken in time $t$, to recognize the relevant estimator $t$, we first need a coarse estimation of the time. Hence, one can introduce coarse estimators of the time through slowly varying part of the dynamics, i.e. the dynamics of the system when is driven with frequency $\omega$. For this purpose, we need only consider the following probabilities:
\begin{gather}\label{pcoarse}
{P_{0+}}(t) = \frac{1}{2}{\cos ^2}\big(\frac{{\omega t}}{2}\big),\\
{P_{0-}}(t) = \frac{1}{2}{\sin ^2}\big(\frac{{\omega t}}{2}\big).
\end{gather}

Suppose that for the $n$ two-qubits system, $k_{1}^{'}$ ($k_{2}^{'}$) of them are found in the post-measurement state $\left| {0} \right\rangle \left| {+} \right\rangle$ ($\left| {0} \right\rangle \left| {-} \right\rangle$). The joint probability associated to these measurement outcomes is obtained as follows
\begin{equation}\label{jp}
\resizebox{.95\hsize}{!}{$
{\textit{P}_{{{\textbf{k}}^{'}}}}(t) = {\left( {\frac{1}{2}} \right)^{k_1^{'} + k_2^{'}}}\left( {\begin{array}{*{20}{c}}
	n\\
	{k_1^{'}}
	\end{array}} \right)\left( {\begin{array}{*{20}{c}}
	{n - k_1^{'}}\\
	{k_2^{'}}
	\end{array}} \right){\cos ^{2k_1^{'}}}(\frac{{\omega t}}{2}){\sin ^{2k_2^{'}}}(\frac{{\omega t}}{2})$,}
\end{equation}
where ${{{\textbf{k}}^{'}}}=({k_1^{'}},{k_2^{'}})$. Considering the likelihood function given in Eq.(\ref{jp}) for $\omega  = 0.5$ and $\Omega  = 1$, once again using the MLM leads to the following estimator for coarse estimating of the time:
\begin{equation}\label{coarseestimator}
{{\hat t}_h}({{\textbf{k}}^{'}}) = 4{\tan ^{ - 1}}\big(\frac{{k_1^{'}}}{{k_2^{'}}}\big).
\end{equation}

The estimator ${{\hat t}_h}({{\textbf{k}}^{'}})$ is used to estimate the time interval in which the fine estimators ${{\hat t}_1}(\textbf{k})$ or ${{\hat t}_3}(\textbf{k})$ belong to. Therefore, by combining the fine tuning and coarse tuning estimators, the operational way to read the precise and long time is as follows:
\begin{equation}\label{finalestimator}
    \hat{t}(\textbf{k}) = 
    \begin{cases} 
        \hat{t}_{1}(\textbf{k}), & 0 \leq \hat{t}_{h}(\textbf{k}^{'}) \leq \pi \\
        \hat{t}_{3}(\textbf{k}), & \pi \leq \hat{t}_{h}(\textbf{k}^{'}) \leq 2\pi
   \end{cases}
\end{equation}

From this perspective, our clock is similar to a pointer clock, where its coarse and fines estimator corresponds to the hours and minutes pointers, respectively. However, there is a subtle difference between them: while in the typical pointer clocks the hours pointer is driven by the minutes pointer, in our quantum clock this is coarse estimator that drives the fine estimator.

\begin{figure}[t]
	\centering
	\includegraphics[scale=0.3]{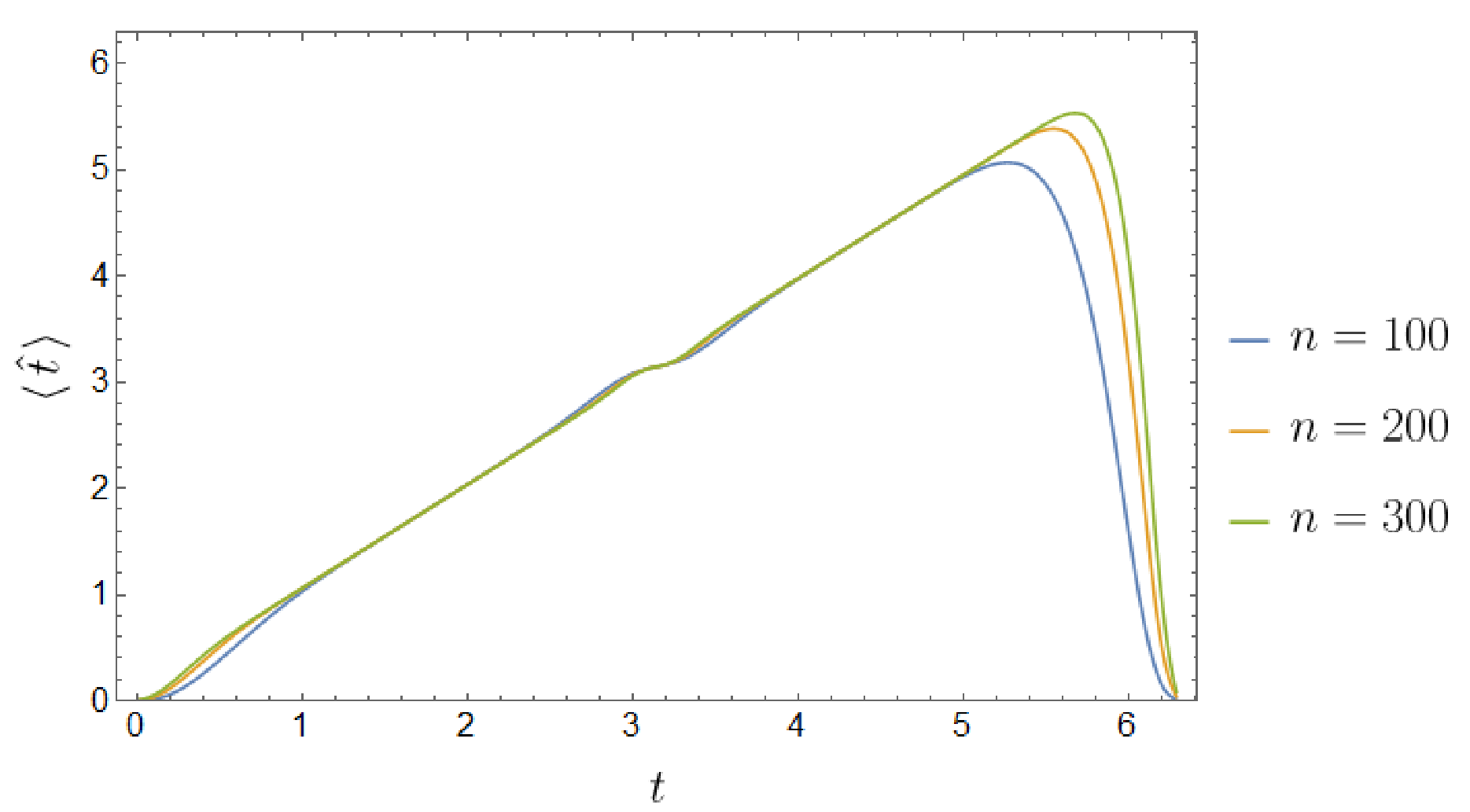}
	\caption{The average of estimator given in Eq.(\ref{finalestimator}) for time interval $[0,2\pi ]$.}\label{GeneralUnbiasedEstimator}
\end{figure}

\begin{figure}[t]
	\centering
	\includegraphics[scale=0.3]{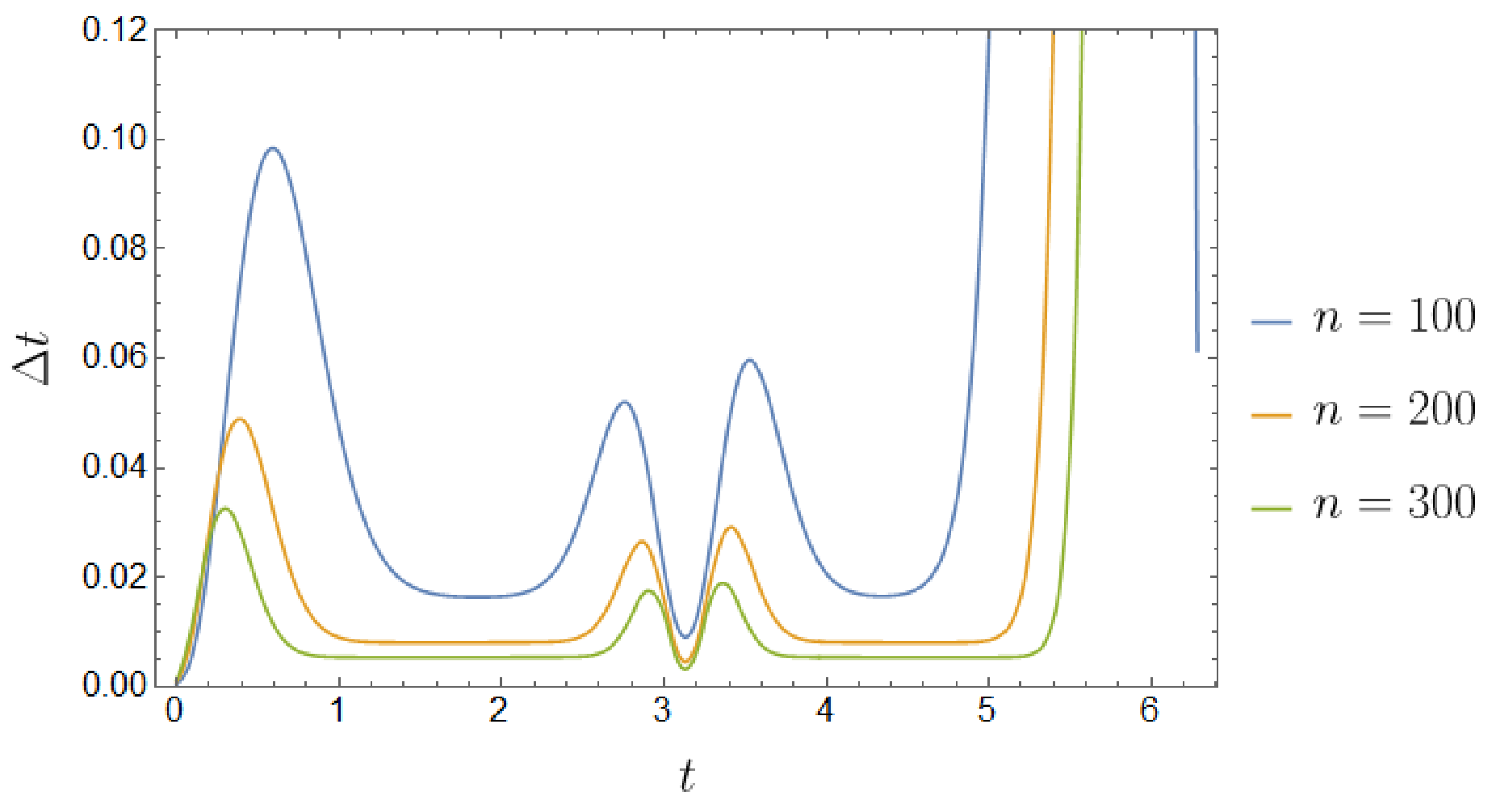}
	\caption{The error of the estimator given in Eq.(\ref{finalestimator}) for time interval $[0,2\pi ]$ and various value of $n$. The horizontal parts of the curves, touch the corresponding Cramer-Rao's lower bound. The green curve touches this bound for longer time compared with the other curves.}\label{varianceofglobalestimator}
\end{figure}

Fig.(\ref{GeneralUnbiasedEstimator}) and Fig.(\ref{varianceofglobalestimator}) depict the average and error of the estimator, respectively. As $n$ grows the estimator becomes more unbiased and it touches the Cramer-Rao's lower bound for longer time.

\section{Quantum Clocks based on Entanglement Resources}
In the previous section we saw that by adding one qubit to the clock system, one can increase the precision and RT of the clock at the same time. In that scenario the time is encoded into the initially separated two-qubits system. However, quantum estimation theory suggests that encoding parameters in an initially entangled state may improve the resolution of the estimators. Therefore it is tempting to examine the characteristics of the quantum clocks with these stronger resources.
To investigate this new scenario, in the first step consider a two-qubits system initially prepared in the following state
\begin{equation}\label{e++}
\left| { +  + } \right\rangle  = \frac{{\left| {00} \right\rangle  + \left| {11} \right\rangle }}{{\sqrt 2 }},
\end{equation}
that is evolved with time according to the following Hamiltonian
\begin{equation}\label{h2}
    H =  - \frac{\omega }{2}I \otimes {\sigma _z} - \frac{\omega }{2}{\sigma _z} \otimes I.
\end{equation}

At some finite time after the evolution of the system, it is subjected to the measurement $M = \left\{ {\left|  \pm  \right\rangle \left\langle  \pm  \right| \otimes \left|  \pm  \right\rangle \left\langle  \pm  \right|} \right\}$. The corresponding output probabilities are as follows
\begin{align}
    \nonumber
    P_{++}(t) &= \frac{1}{2} \cos^2(\omega t),
    &
    P_{+-}(t) &= \frac{1}{2} \sin^2(\omega t),
    \\
    \label{M2}
    P_{-+}(t) &= \frac{1}{2} \sin^2\big(\omega t),
    &
    P_{--}(t) &= \frac{1}{2} \cos^2(\omega t).
\end{align}

It is a straightforward task to show the Fisher information of this probability distribution is $4{\omega ^2}$. Therefore, using the Cramer-Rao inequality for a fixed number of $n/2$ samples, the resolution of the time-estimator is bounded by the quantity $1/\sqrt{(n/2)\left( {4{\omega ^2}} \right)} = 1/\sqrt{2n{\omega ^2}}$. In the second step consider the number of $n/3$ samples of 3-qubits systems initially prepared in the state $\left| { +  +  + } \right\rangle  = (\left| {000} \right\rangle  + \left| {111} \right\rangle )/\sqrt 2 $ and evolved with Hamiltonian 
\begin{equation}\label{h3}
H =  - \frac{\omega }{2}I \otimes I \otimes {\sigma _z} - \frac{\omega }{2}I \otimes {\sigma _z} \otimes I - \frac{\omega }{2}{\sigma _z} \otimes I \otimes I,
\end{equation}
and consequently subjected to the measurement $ M = \left\{ {\left|  \pm  \right\rangle \left\langle  \pm  \right| \otimes \left|  \pm  \right\rangle \left\langle  \pm  \right| \otimes \left|  \pm  \right\rangle \left\langle  \pm  \right|} \right\}$. The corresponding output probabilities are as follows
\begin{align}
    \nonumber
    P_{+++}(t) &= \frac{1}{4} \cos^2\big(\frac{3\omega t}{2}\big),
    &
    P_{++-}(t) &= \frac{1}{4} \sin^2\big(\frac{3\omega t}{2}\big),
    \\
    \nonumber
    P_{+-+}(t) &= \frac{1}{4} \sin^2\big(\frac{3\omega t}{2}\big),
    &
    P_{+--}(t) &= \frac{1}{4} \cos^2\big(\frac{3\omega t}{2}\big),
    \\
    \nonumber
    P_{-++}(t) &= \frac{1}{4} \cos^2\big(\frac{3\omega t}{2}\big),
    &
    P_{-+-}(t) &= \frac{1}{4} \sin^2\big(\frac{3\omega t}{2}\big),
    \\
    \label{3prob}
    P_{--+}(t) &= \frac{1}{4} \sin^2\big(\frac{3\omega t}{2}\big),
    &
    P_{---}(t) &= \frac{1}{4} \cos^2\big(\frac{3\omega t}{2}\big).
\end{align}

\begin{figure}[t]
	\centering
	\includegraphics[scale=0.4]{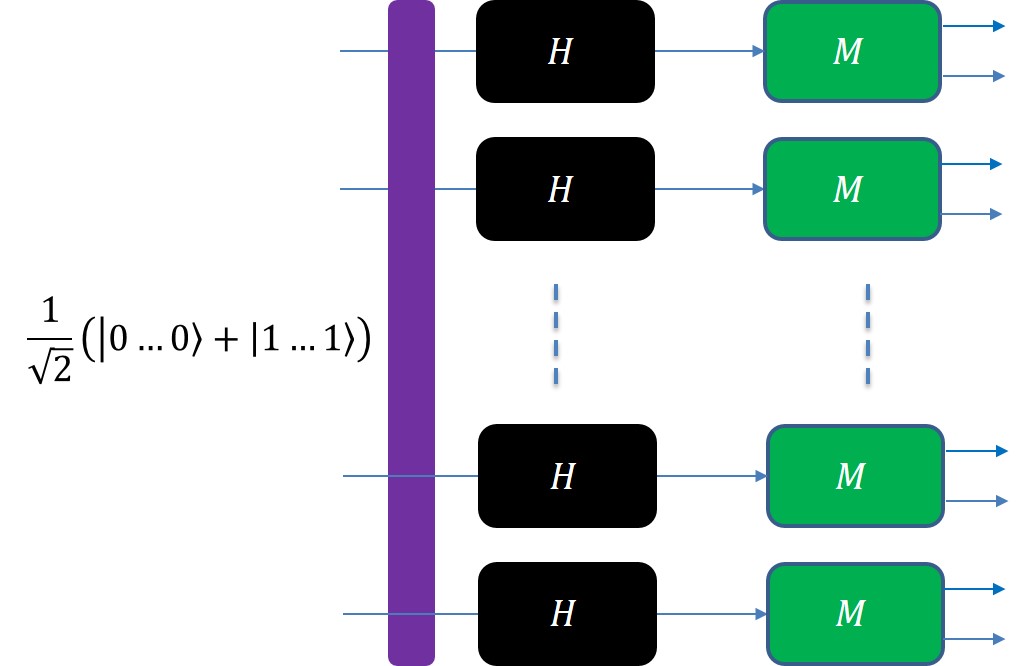}
	\caption{Schematic model of quantum clock based on dynamics of $n$ entangled qubits. The Entangled $n$-qubits system is initially prepared in the  state ${{(\left| {0...0} \right\rangle  + \left| {1...1} \right\rangle )} \mathord{\left/
 {\vphantom {{(\left| {0...0} \right\rangle  + \left| {1...1} \right\rangle )} {\sqrt 2 }}} \right.
 \kern-\nulldelimiterspace} {\sqrt 2 }}$ and then evolves with the Hamiltonian given in Eq.(\ref{h3}). At finite time $t$ a measurement in the basis $\left\{ {\left|  \pm  \right\rangle ...\left|  \pm  \right\rangle } \right\}$ is carried out on the system. From the statistical of the measurement outputs the time $t$ can be estimated, where the minimum error of the estimation scales as $1/n\omega $. Although the model based on entanglement resource improves the precision of quantum clocks, it cannot support long RT (see the text).}\label{scenario3}
\end{figure}

Similar calculations show that in this case the resolution of the time-estimator is bounded by the quantity $1/\sqrt{(n/3)\left( {9{\omega ^2}} \right)} = 1/\sqrt{3n{\omega ^2}}$ as well. Consequently, one easily derives that the resolution of the time estimator for $n$ entangled samples is bounded by the quantity $1/\sqrt{{n^2}{\omega ^2}}$ that is nothing but a well-known bound called Heisenberg limit. Therefore, by exploiting more strong resources, i.e. the number of $n$ entanglement resources, the resolution of the estimators scales as $1/{n}$ that improves the Cramer-Rao bound by a factor of $1/\sqrt{n}$.  However, a careful comparison of Eq.(\ref{jointprob}) and Eq.(\ref{M2}) recovers the fact that the increase of the precision of the estimator is solely due to the increase of the evolution frequency of the system (for $n=2$ evolution frequency of the entangled system becomes twice the evolution frequency of separable system), that in turns reduces the RT of the clock. Thus, while entanglement improves the precision of the quantum clock, it inevitably worsens the RT of the clock, i.e. there exists a intrinsic tradeoff between the two quantities. Fig.(\ref{scenario3}) represents the schematic model of the quantum clock based on $n$ entangled qubits.

\section{Conclusion}
Continuous clocks are regarded as essential components of sensing technology. In addition to precision, RT is one of the main features of this kind of clocks. Therefore, proposing and characterizing a continuous clock with optimal precision and long RT is of great interest. In this paper, we used the approach of quantum estimation theory to propose a continuous quantum clock with optimal characteristics, i.e. high precision and long RT. In this context, we investigated the behavior of the quantum clock with mentioned characteristics by employing various tools of QET. 

In the first proposal, the time is encoded into the dynamics of a one-qubit system via an optimal Hamiltonian and subsequently decoded via an optimal projective measurement. In this scenario, the resolution of the optimal time estimator for the number of $n$ qubits scales as $1/\sqrt n$. But for a fixed amount of resources, the precision and RT of the clock are inversely proportional i.e.  this scenario cannot support long RT and high precision at the same time. To deal with this challenge, the dynamics of coherent two-qubits system was introduced as second proposal of quantum clock, where the frequency of the slowly varying part of the dynamics determines the RT of the clock and can be set to arbitrary small values corresponding to large values of the RT. Instead, this is the frequency of the rapidly varying part of the dynamics that determines the precision of the quantum clock and independent of the former frequency can be set to arbitrary large values. In this way, the characteristics of a quantum clock with high precision and long RT was obtained. In this scenario, the precision of the clock scales as $1/\sqrt n$ as well. Therefore, at first glance one may think that apart from the RT it doesn’t make a better precision than first scenario. However, a true comparison based on the number of used resource states shows that the second scenario (QC based on $n$ two-qubits system) can do better than first scenario ($2n$ one-qubit system). As an example, one can check that for values $\Omega  = 1.0$ and $\omega  = 0.5$ introduced in Sec.\ref{schenario2section}, the quantum clock based on $n$ two-qubits system has better precision than quantum clock based on $2n$ one-qubit system. 

Quantum mechanics provide us with more powerful resource, i.e. entanglement resource, which may improve the performance of the clocks. It was shown, using $n$ entanglement resources, improves the precision of the quantum clock by a factor of $1/\sqrt n$. However, in this case the increase of the precision is solely due to the increase of the evolution frequency of the system that in turns reduces the RT of the clock. Thus, while entanglement improves the precision of the quantum clock, it inevitably worsens the RT of the clock, i.e. there exists a intrinsic tradeoff between the two quantities. 

\bibliography{quantum_clock.bib}

\end{document}